\documentclass[12pt]{iopart}

\usepackage{iopams}
\usepackage{verbatim} 
\usepackage[utf8]{inputenc}

\usepackage{graphicx}
\usepackage[colorlinks=true,citecolor=blue]{hyperref}
\usepackage{units}

\newcommand{\text}[1]{\mathsf{#1}}

\def\kB {k_\mathsf{B}}
\newcommand{\kBT}{k_\mathsf{B}T}

\begin{document}

\title{Powerful energy harvester based on resonant-tunneling quantum wells}

\author{Bj\"orn Sothmann$^1$, Rafael Sánchez$^2$, Andrew N. Jordan$^3$ and Markus B\"uttiker$^1$}
\address{$^1$D\'epartement de Physique Th\'eorique, Universit\'e de Gen\`eve, CH-1211 Gen\`eve 4, Switzerland}
\address{$^2$Instituto de Ciencia de Materiales de Madrid (ICMM-CSIC), Cantoblanco, E-28049 Madrid, Spain}
\address{$^3$Department of Physics and Astronomy, University of Rochester, Rochester, New York 14627, USA}
\eads{\mailto{bjorn.sothmann@unige.ch},\\
\mailto{rafael.sanchez@icmm.csic.es},\\
\mailto{jordan@pas.rochester.edu},\\
\mailto{markus.buttiker@unige.ch}}

\date{\today}

\begin{abstract}
We analyze a heat engine based on a hot cavity connected via quantum wells to electronic reservoirs. We discuss the output power as well as the efficiency both in the linear and nonlinear regime. We find that the device delivers a large power of about $\unit[0.18]{W/cm^2}$ for a temperature difference of $\unit[1]{K}$ nearly doubling the power than can be extracted from a similar heat engine based on quantum dots. At the same time, the heat engine also has a good efficiency although reduced from the quantum dot case. Due to the large level spacings that can be achieved in quantum wells, our proposal opens the route towards room-temperature applications of nanoscale heat engines.
\end{abstract}


\submitto{\NJP}

\maketitle

\section{\label{sec:introduction}Introduction}
Energy harvesters collect energy from the environment and use it to power small electronic devices or sensors~\cite{white_energy-harvesting_2008}. By now, a wide variety of energy harvesters have been proposed that convert ambient energy to electrical or mechanical power, e.g., from vibrations, electromagnetic radiation or by relying on thermoelectric effects. The latter systems turn out to be particularly useful to convert heat on a computer chip back into electrical power, thereby reducing both the power consumption of the chip as well as the need to actively cool it.

The main challenge of current research on thermoelectric energy harvesters is to find setups that are both powerful and efficient at the same time. Artificially fabricated nanoscale structures are promising candidates for highly efficient thermoelectrics. Already 20 years ago, Hicks and Dresselhaus demonstrated that mesoscopic one-dimensional wires~\cite{hicks_thermoelectric_1993} as well as quantum wells~\cite{hicks_effect_1993} have thermoelectric figures of merit greatly enhanced compared to the bulk values. Mahan and Sofo later showed that the best thermoelectric properties occur in materials that are good energy filters, i.e., have sharp spectral features~\cite{mahan_best_1996}.

A paradigmatic realization of such spectral features is given by quantum dots with sharp, discrete energy levels. The thermopower of quantum dots in the Coulomb-blockade regime coupled to two electronic reservoirs at different temperatures has been studied both theoretically~\cite{beenakker_theory_1992} and experimentally~\cite{staring_coulomb-blockade_1993,molenkamp_sawtooth-like_1994,dzurak_thermoelectric_1997,scheibner_thermopower_2005,scheibner_sequential_2007,svensson_lineshape_2012}. Later, the thermopower of open quantum dots~\cite{godijn_thermopower_1999} and carbon nanotube quantum dots~\cite{llaguno_observation_2003,small_modulation_2003} was investigated. Thermoelectric effects have also been studied for resonant tunneling through a single quantum dot~\cite{nakpathomkun_thermoelectric_2010}. Compared to a weakly-coupled quantum dot in the Coulomb-blockade regime~\cite{esposito_thermoelectric_2009}, the power is enhanced while at the same time the efficiency is reduced due to the finite level width of the resonant state. More complicated resonant tunneling configurations have been proposed to optimize the efficiency at finite power output~\cite{whitney_best_2013,hershfield_non-linear_2013}.

Recently, there has been a growing interest in thermoelectrics with three-terminal structures~\cite{sanchez_optimal_2011,sothmann_rectification_2012,jordan_powerful_2013,sothmann_magnon-driven_2012,entin-wohlman_three-terminal_2010,entin-wohlman_three-terminal_2012,jiang_thermoelectric_2012,ruokola_theory_2012,sanchez_detection_2012,jiang_three-terminal_2013,bergenfeldt_hybrid_2013}. Four-terminal configurations (two Coulomb-coupled conductors subject to currents) have been of interest in the discussion of nonequilibrium fluctuations~\cite{sanchez_mesoscopic_2010,schaller_low-dimensional_2010,krause_incomplete_2011,bulnes_cuetara_fluctuation_2011,golubev_fluctuation_2011}. In addition, reciprocity relations for multi-terminal thermoelectric transport have been analysed~\cite{butcher_thermal_1990,sanchez_thermoelectric_2011,jacquod_onsager_2012,matthews_experimental_2013,hwang_magnetic-field_2013}. Compared to conventional two-terminal setups, they offer the advantage of separating the heat and charge current flow. Furthermore, they naturally operate in a conventional thermocouple-like fashion.
While a system of two Coulomb-coupled quantum dots in the Coulomb-blockade regime was shown to work as an optimal heat-to-current converter that can reach Carnot efficiency, the resulting currents and output powers are limited by the fact that transport only proceeds via tunneling of single electrons~\cite{sanchez_optimal_2011}. A related setup where the Coulomb-blockade dots are replaced by chaotic cavities connected via quantum point contacts with a large number of open transport channels to the electronic reservoirs turns out to deliver much larger currents. Nevertheless, the resulting output power is similar to the Coulomb-blockade setup since the thermoelectric performance is determined by the weak energy dependence of a single partially open transport channel~\cite{sothmann_rectification_2012}. This problem can be overcome by a heat engine based on resonant-tunneling quantum dots. Such a system yields a large output power of $\unit[0.1]{pW}$ for a temperature difference of $\unit[1]{K}$ between the hot and the cold reservoir while at the same time it reaches an efficiency at maximum power of about 20\% of the Carnot efficiency. In addition, the device can be scaled to macroscopic dimensions by parallelization based on the use of self-assembled quantum dots~\cite{jordan_powerful_2013}.
Similar setups have also been investigated both theoretically~\cite{edwards_quantum-dot_1993,edwards_cryogenic_1995} and experimentally~\cite{prance_electronic_2009} in their dual rôle as refrigerators.

Here, we analyze the performance of a three-terminal energy harvester based on resonant quantum wells. Our work is motivated by a number of advantages that we expect a quantum-well structure to have over a quantum-dot setup. First of all, quantum wells should be able to deliver larger currents and, therefore, larger output powers because of the transverse degrees of freedom. The available phase space for electrons that can traverse the well is large. Second, a quantum-well structure might be easier to fabricate than a system of self-assembled quantum dots that all should have similar properties in order to yield a decent device performance although there is good tolerance to fluctuations in dot properties~\cite{jordan_powerful_2013}. Finally, due to the large level spacing of narrow quantum wells, they are ideally suited for room-temperature applications. Apart from these advantages, we also aim to investigate how the less optimal energy-filtering properties of quantum wells compared to quantum dots deteriorate the efficiency of heat-to-current conversion (quantum wells transmit any electron with an energy larger than the level position whereas quantum dots transmit only electrons with an energy exactly equal to the level energy).

Our paper is organized as follows. In section~\ref{sec:setup}, we introduce the model of the quantum-well harvester. We then present our results for the power and efficiency in section~\ref{sec:results} analyzing both the linear and the nonlinear transport regime. We finally give some conclusions in section~\ref{sec:conclusions}.

\section{\label{sec:setup}Setup}
\begin{figure}
	\centering\includegraphics[width=.8\textwidth]{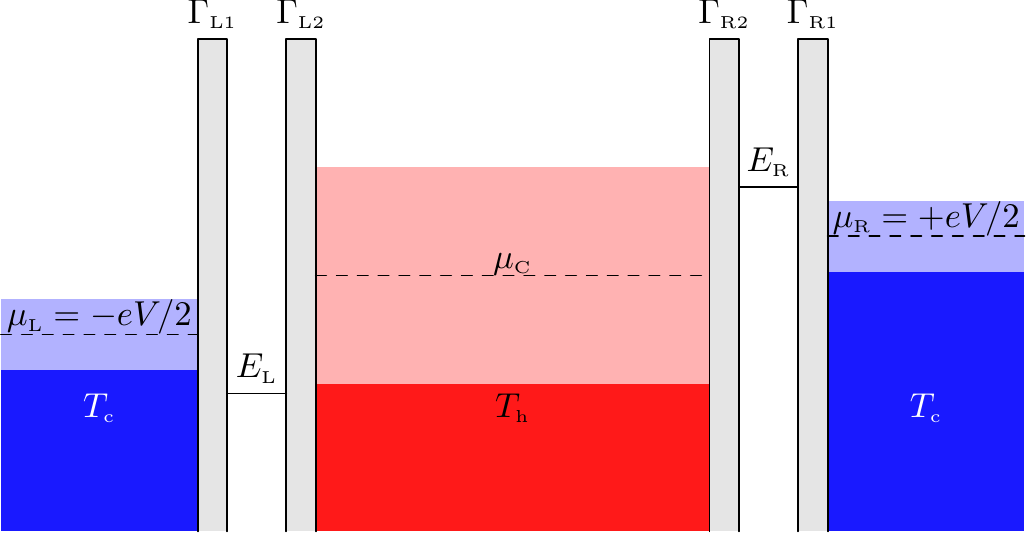}
	\caption{\label{fig:model}Schematic of the quantum-well based energy harvester. A central cavity (red) kept at temperature $T_\text{h}$ by a hot thermal reservoir (not shown) is connected via quantum wells to two electron reservoirs at temperature $T_\text{c}$ (blue). Chemical potentials are measured relative to the equilibrium chemical potential.}
\end{figure}
The system we consider is schematically shown in \fref{fig:model}. It consists of a central cavity connected via quantum wells to two electronic reservoirs. In the following, we assume the quantum wells to be noninteracting such that charging effects can be neglected in a simplified model. We will revisit the effects of interactions relevant in the nonlinear regime with a more realistic treatment in the future.

The electronic reservoirs, $r=\text{L,R}$, are characterized by a Fermi function $f_r(E)=\{\exp[(E-\mu_r)/(\kB T_\text{c})]+1\}^{-1}$ with temperature $T_\text{c}$ and chemical potentials $\mu_r$. 
The cavity is assumed to be in thermal equilibrium with a heat bath of temperature $T_\text{h}$. The nature of this heat bath is not relevant for our discussion and depends on the source from which we want to harvest energy. Strong electron-phonon and electron-electron interactions within the cavity relax the energy of the electrons entering and leaving the cavity towards a Fermi distribution $f_\text{C}(E)=\{\exp[(E-\mu_\text{C})/(\kB T_\text{h})]+1\}^{-1}$ characterized by the cavity temperature $T_\text{h}$ and the cavity's chemical potential $\mu_\text{C}$.

The cavity potential $\mu_\text{C}$ as well as its temperature $T_\text{h}$ (or, equivalently, the heat current $J$ injected from the heat bath into the cavity to keep it at a given temperature $T_\text{h}$) have to be determined from the conservation of charge and energy, $I_\text{L}+I_\text{R}=0$ and $J^\text{E}_\text{L}+J^\text{E}_\text{R}+J=0$. Here, $I_r$ denotes the current flowing from reservoir $r$ into the cavity. Similarly, $J^\text{E}_r$ denotes the energy current flowing from reservoir $r$ into the cavity.

The charge and energy currents can be evaluated within a scattering matrix approach as~\cite{blanter_transition_1999} 
\begin{equation}\label{eq:I}
	I_r=\frac{e\nu_2\mathcal A}{2\pi\hbar}\int dE_\perp dE_z T_r(E_z)\left[f_r(E_z+E_\perp)-f_\text{C}(E_z+E_\perp)\right],
\end{equation}
and
\begin{equation}\label{eq:JE}
	J^\text{E}_{r}=\frac{\nu_2\mathcal A}{2\pi\hbar}\int dE_\perp dE_z (E_z+E_\perp)T_r(E_z)\left[f_r(E_z+E_\perp)-f_\text{C}(E_z+E_\perp)\right].
\end{equation}
Here, $\nu_2=m_*/(\pi\hbar^2)$ is the density of states of the two-dimensional electron gas inside the quantum well with the effective electron mass $m_*$. $\mathcal A$ denotes the surface area of the well. $E_z$ and $E_\perp$ are the energy associated with motion in the well's plane and perpendicular to it, respectively.
The transmission of quantum well $r$ is given by~\cite{buttiker_coherent_1988}
\begin{equation}
	T_r(E)=\frac{\Gamma_{r1}(E)\Gamma_{r2}(E)}{(E-E_{nr})^2+[\Gamma_{r1}(E)+\Gamma_{r2}(E)]^2/4}.
\end{equation}
Here, $\Gamma_{r1}(E)$ and $\Gamma_{r2}(E)$ denote the (energy-dependent~\cite{blanter_transition_1999}) coupling strength of the quantum well to the electronic reservoir $r$ and the cavity, respectively. The energies of the resonant levels (more precisely the subband thresholds) within the quantum well are given by $E_{nr}$. For a parallel geometry with well width $L$, the resonant levels are simply given by the discrete eigenenergies of a particle in a box, $E_{nr}=(\pi\hbar n)^2/(2m^* L^2)$. In the following, we always restrict ourselves to the situation of weak couplings, $\Gamma_{r1},\Gamma_{r2}\ll \kB T_\text{c},\kB T_\text{h}$, whose energy dependence can be neglected. Furthermore, we assume that the level spacing inside the quantum wells is large such that only the lowest energy state is relevant for transport. In this case, the transmission function reduces to a single delta peak, $T_r(E)=2\pi\Gamma_{1r}\Gamma_{2r}(\Gamma_{1r}+\Gamma_{2r})\delta(E_z-E_{1r})$. This allows us to analytically solve the integrals in the expressions \eref{eq:I} and \eref{eq:JE} for the currents and yields
\begin{equation}
	I_r=\frac{e\nu_2\mathcal A}{\hbar}\frac{\Gamma_{r1}\Gamma_{r2}}{\Gamma_{r1}+\Gamma_{r2}}\left[\kB T_\text{c}K_1\left(\frac{\mu_r-E_r}{\kB T_\text{c}}\right)-\kB T_\text{h} K_1\left(\frac{\mu_\text{C}-E_r}{\kB T_\text{h}}\right)\right],
\end{equation}
as well as
\begin{equation}
	\fl J^\text{E}_r=\frac{E_{r}}{e}I_r
	+\frac{\nu_2\mathcal A}{\hbar}\frac{\Gamma_{r1}\Gamma_{r2}}{\Gamma_{r1}+\Gamma_{r2}}\left[(\kB T_\text{c})^2K_2\left(\frac{\mu_r-E_r}{\kB T_\text{c}}\right)-(\kB T_\text{h})^2K_2\left(\frac{\mu_\text{C}-E_r}{\kB T_\text{h}}\right)\right],
\end{equation}
where for simplicity we denote the energy of the single resonant level in the quantum wells as $E_r$. We furthermore introduced the integrals $K_1(x)=\int_0^\infty \rmd t(1+e^{t-x})^{-1}=\log(1+e^x)$ and $K_2(x)=\int_0^\infty \rmd t\; t(1+e^{t-x})^{-1}=-{\rm Li}_2(-e^x)$ with the dilogarithm ${\rm Li}_2(z)=\sum_{k=1}^\infty \frac{z^k}{k^2}$.
The heat current is made up from two different contributions. While the first one is simply proportional to the charge current, the second term breaks this proportionality. We remark that in the case of quantum dots with sharp levels, the latter term is absent~\cite{jordan_powerful_2013}.

\section{\label{sec:results}Results}
In the following, we first analyse the system in the linear-response regime and then turn to the nonlinear situation. We assume that both quantum wells are intrinsically symmetric, i.e., $\Gamma_{\text{L}1}=\Gamma_{\text{L}2}\equiv(1+a)\Gamma$, $\Gamma_{\text{R}1}=\Gamma_{\text{R}2}\equiv(1-a)\Gamma$. Here, $\Gamma$ denotes the total coupling strength whereas $-1\leq a\leq 1$ characterizes the asymmetry between the coupling of the left and the right well.

\subsection{\label{ssec:linear}Linear response}
We start our analysis by a discussion of the linear-response regime. To simplify notation, we introduce the average temperature $T=(T_\text{h}+T_\text{c})/2$ and the temperature difference $\Delta T=T_\text{h}-T_\text{c}$. To linear order in the temperature difference $\Delta T$ and the bias voltage $eV=\mu_\text{R}-\mu_\text{L}$ applied between the two electronic reservoirs, the charge current through the system is given by
\begin{equation}
	I_\text{L}=-I_\text{R}=\frac{e\nu_2\mathcal A\Gamma}{2\hbar}g_1\left(\frac{E_\text{L}}{\kB T},\frac{E_\text{R}}{\kB T}\right)\left[-eV-\kB\Delta T g_2\left(\frac{E_\text{L}}{\kB T},\frac{E_\text{R}}{\kB T}\right)\right],
\end{equation}
with the auxiliary functions
\begin{equation}
	g_1(x,y)=\frac{1-a^2}{2+(1-a)e^x+(1+a)e^y},
\end{equation}
and
\begin{equation}
	g_2(x,y)=x-y+(1+e^x)\log(1+e^{-x})-(1+e^y)\log(1+e^{-y}).
\end{equation}
At $V=0$, a finite current driven by $\Delta T\neq0$ flows in a direction that depends on the position of the resonant levels. If, e.g., $E_\text{R}>E_\text{L}$, electrons will be transferred from the left to the right lead

The power delivered by the heat-driven current against the externally applied bias voltage $eV$ is simply given by $P=I_\text{L}V$. It vanishes at zero applied voltage. Furthermore, it also vanishes at the so called stopping voltage $V_\text{stop}$ where the heat-driven current is exactly compensated by the bias-driven current flowing in the opposite direction. In between these two extreme cases, the output power depends quadratically on the bias voltage and takes its maximal value at half the stopping voltage. Here, the maximal output power is given by
\begin{equation}
	P_\text{max}=\frac{\nu_2\mathcal A\Gamma}{2\hbar}\left(\frac{\kB\Delta T}{2}\right)^2 g_1\left(\frac{E_\text{L}}{\kB T},\frac{E_\text{R}}{\kB T}\right)g_2^2\left(\frac{E_\text{L}}{\kB T},\frac{E_\text{R}}{\kB T}\right).
\end{equation}
The effiency $\eta$ of the quantum-well heat engine is defined as the ratio between the output power and the input heat. The latter is given by the heat current $J$ injected from the heat bath, i.e., we have $\eta=P/J$.
For a bias voltage $V=V_\text{stop}/2$ that delivers the maximal output power, the heat current is given by
\begin{equation}\label{eq:J}
	J=\frac{\nu_2\mathcal A\Gamma}{2\hbar}(\kB T)^2 \frac{\Delta T}{T} g_3\left(\frac{E_\text{L}}{\kB T},\frac{E_\text{R}}{\kB T}\right),
\end{equation}
where the function $g_3(x,y)$ that satisfies $0<g_3(x,y)<2\pi^2/3$ is given in the Appendix for completeness.
Hence, the efficiency at maximum power is simply given by
\begin{equation}
	\eta_\text{maxP}=\frac{\eta_\text{C}}{4}\frac{g_1\left(\frac{E_\text{L}}{\kB T},\frac{E_\text{R}}{\kB T}\right)g_2^2\left(\frac{E_\text{L}}{\kB T},\frac{E_\text{R}}{\kB T}\right)}{g_3\left(\frac{E_\text{L}}{\kB T},\frac{E_\text{R}}{\kB T}\right)},
\end{equation}
with the Carnot efficiency $\eta_\text{C}=1-\frac{T_\text{c}}{T_\text{h}}\approx\frac{\Delta T}{T}$.

\begin{figure}
	\includegraphics[width=.49\textwidth]{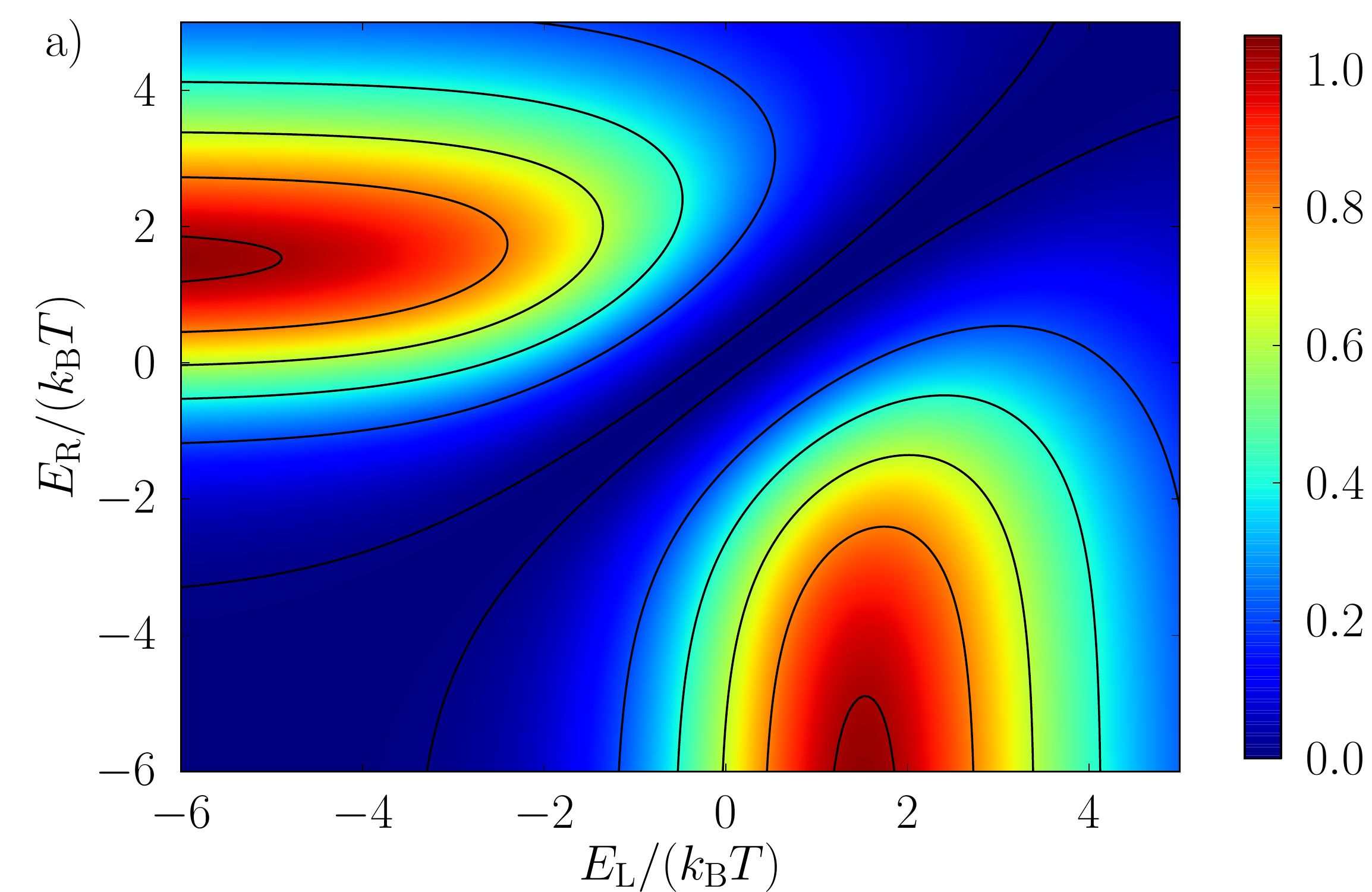}
	\includegraphics[width=.49\textwidth]{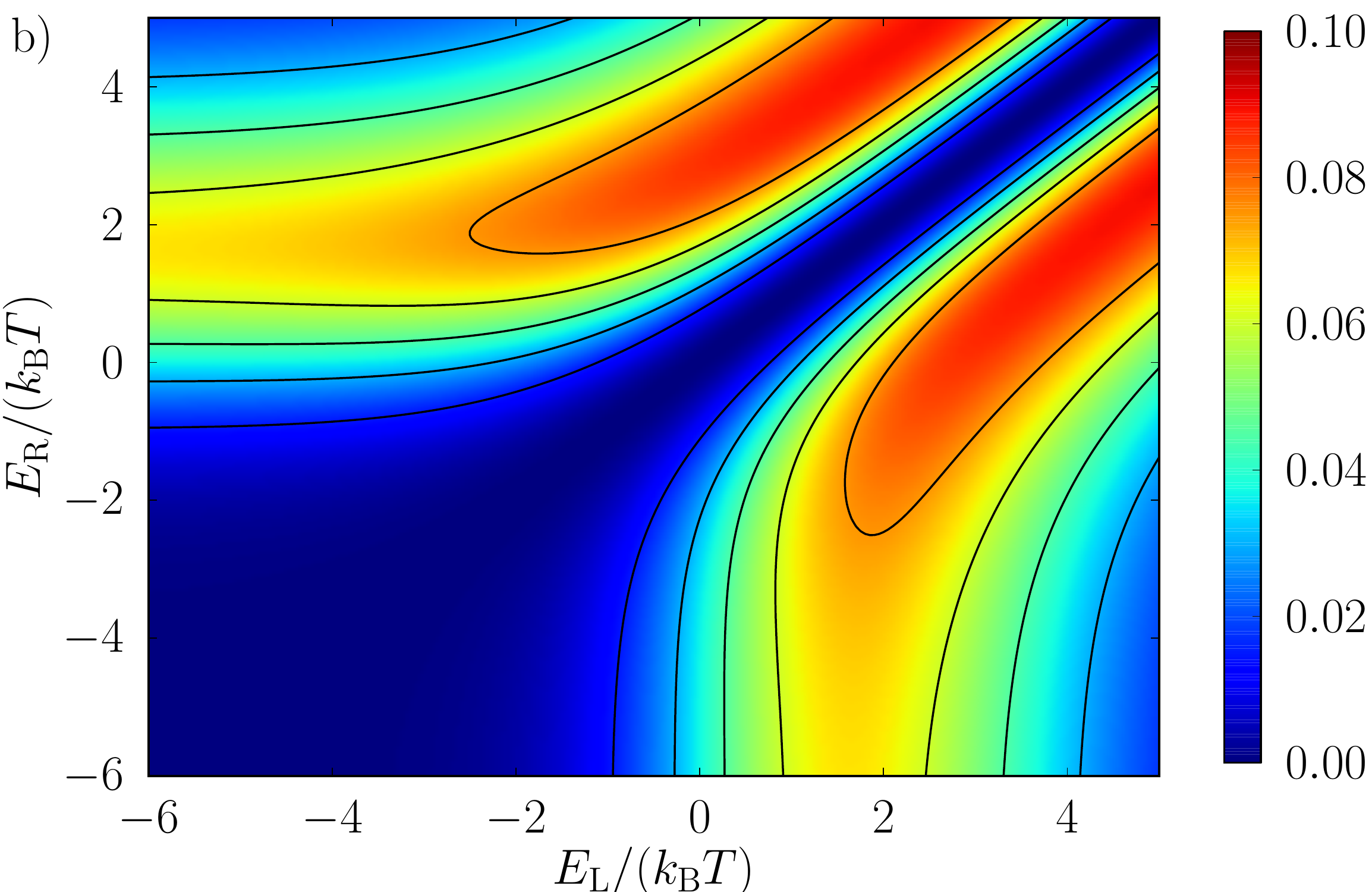}
	\includegraphics[width=.49\textwidth]{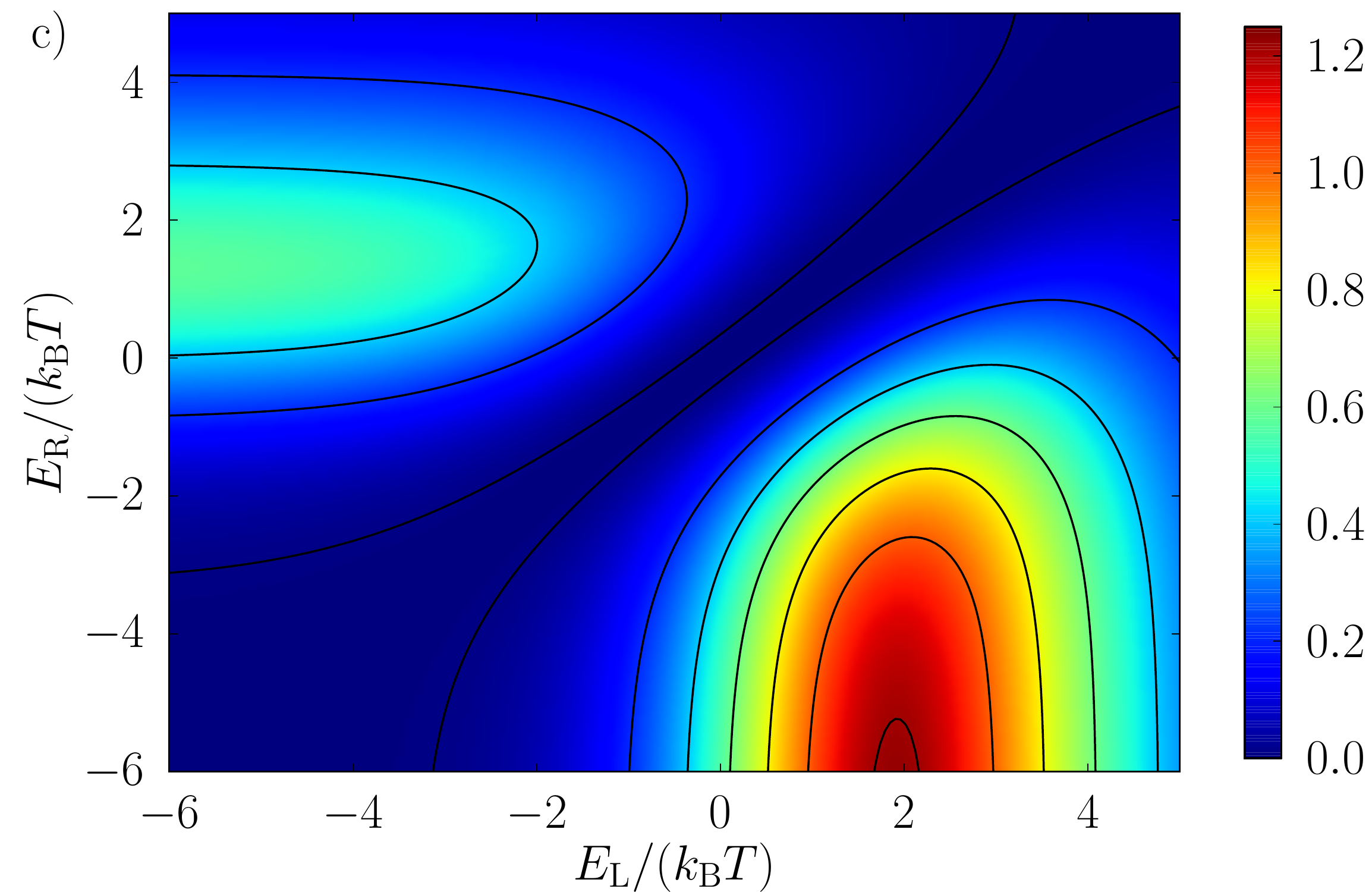}
	\includegraphics[width=.49\textwidth]{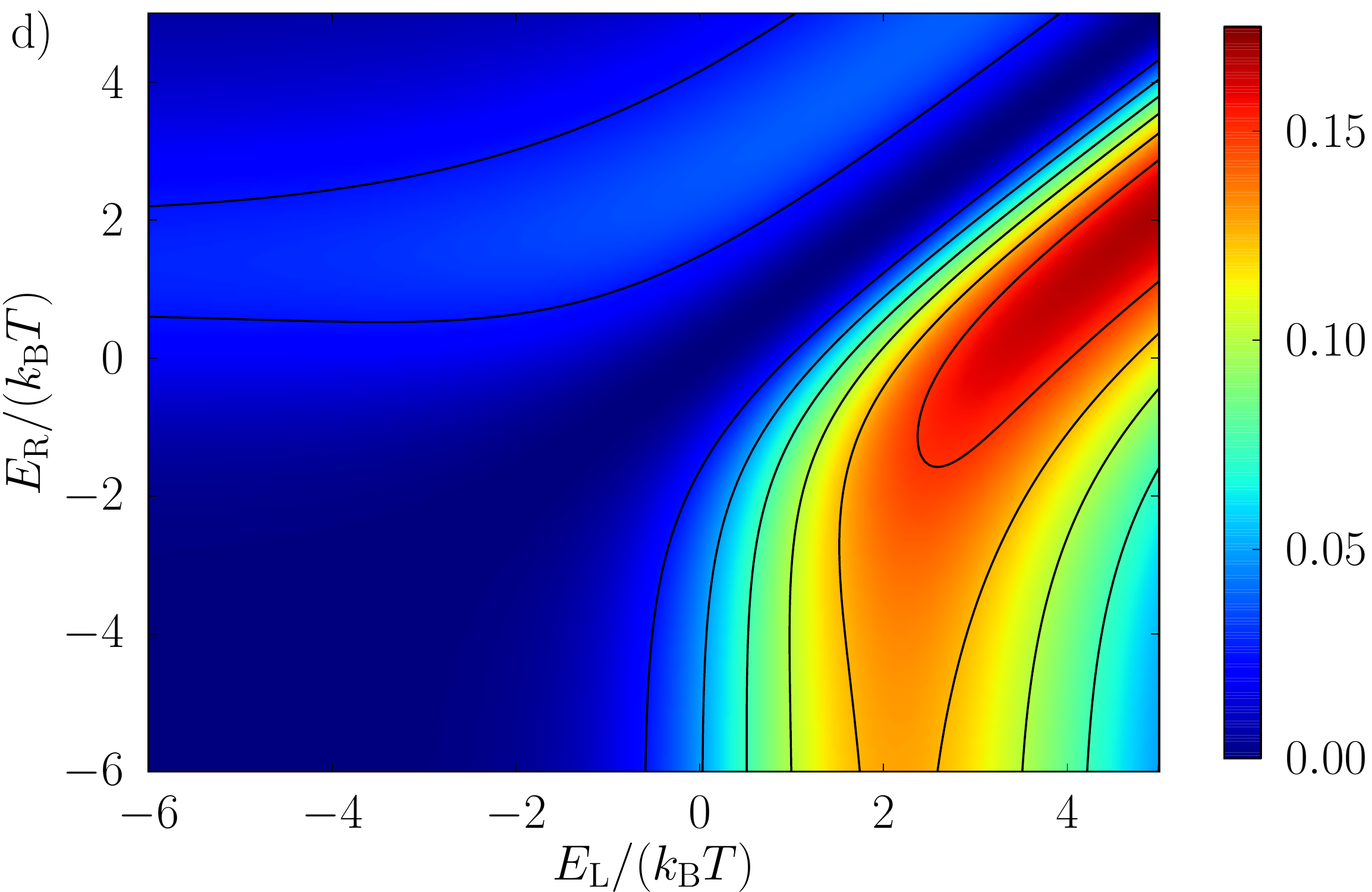}
	\caption{\label{fig:Linear}a) Maximum power in units of $\frac{\nu_2\mathcal A\Gamma}{2\hbar}\left(\frac{\kB\Delta T}{2}\right)^2$ within linear response as a function of the level positions inside the two quantum wells for a symmetric setup $a=0$. b) Efficiency at maximum power in units of the Carnot efficiency $\eta_\text{C}$ within linear response as a function of the two level positions for a symmetric configuration. c) and d) show the same as a) and b) but for a system with asymmetry $a=0.5$.}
\end{figure}
\begin{figure}
	\includegraphics[width=.49\textwidth]{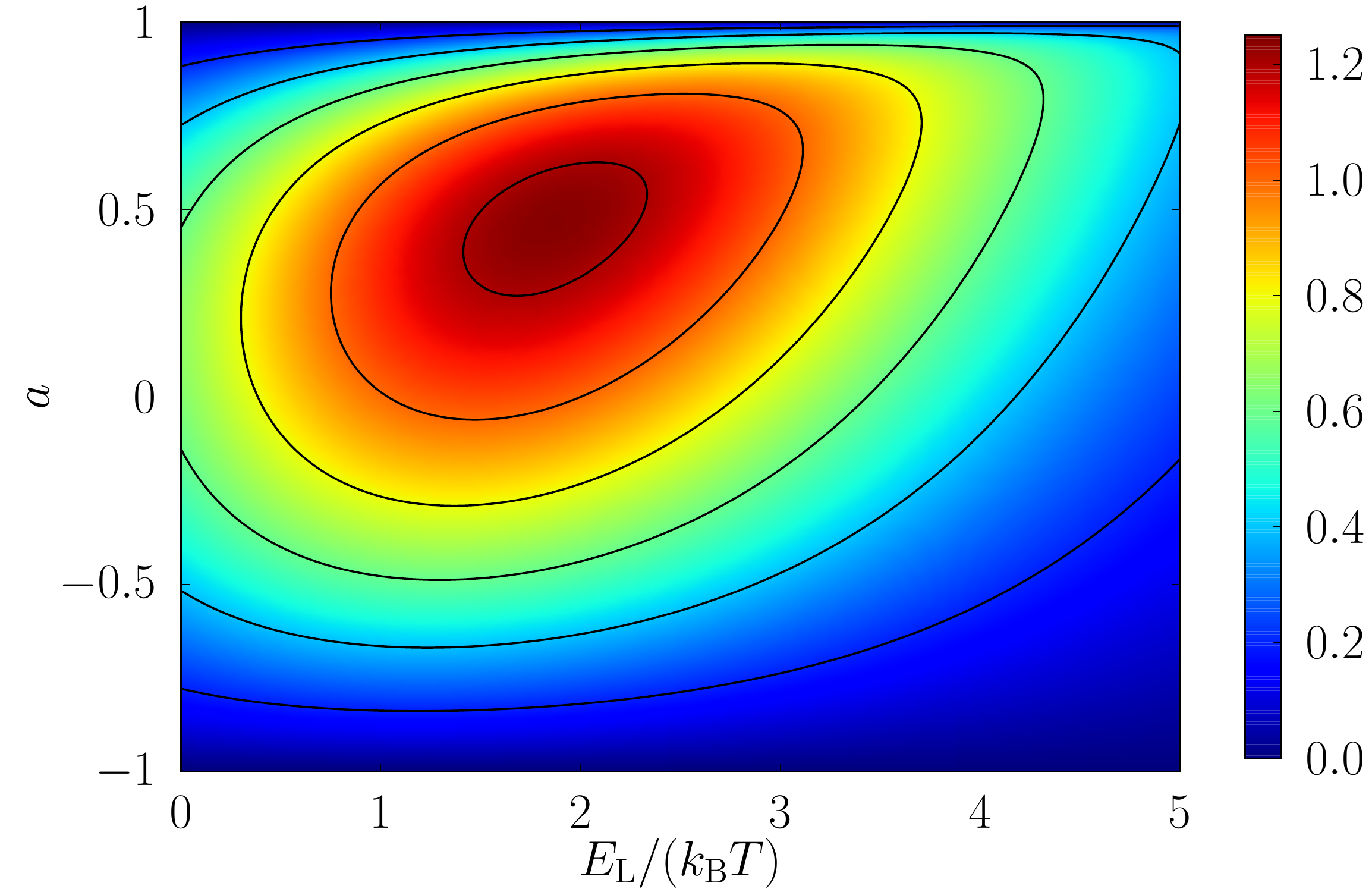}
	\includegraphics[width=.49\textwidth]{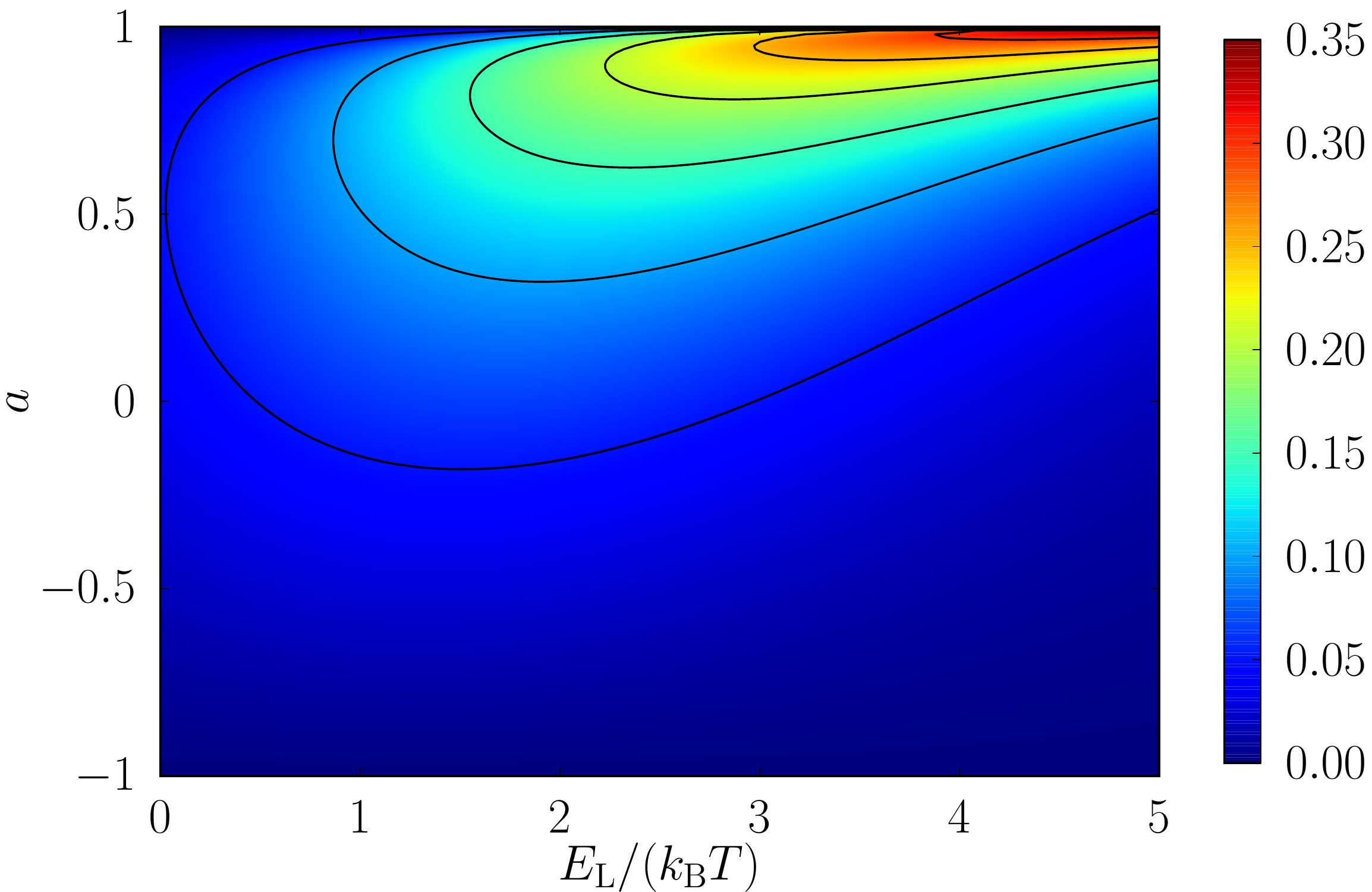}
	\caption{\label{fig:LinearAsymmetry}Left panel: Maximum power in units of $\frac{\nu_2\mathcal A\Gamma}{2\hbar}\left(\frac{\kB\Delta T}{2}\right)^2$ within linear response as a function of one level position and the asymmetry of couplings. Right panel: Efficiency at maximum power in units of the Carnot efficiency $\eta_\text{C}$ within linear response as a function of one level position and the asymmetry of couplings. For both plots, $E_\text{R}=-10\kBT$.}
\end{figure}

We now discuss the output power and the efficiency in more detail, first focussing on a symmetric system, $a=0$. In \fref{fig:Linear}, we show the power as a function of the level positions $E_\text{L}$ and $E_\text{R}$. It is symmetric with respect to an exchange of $E_\text{L}$ and $E_\text{R}$. The maximal output power of approximately $P_\text{max}\approx\frac{\nu_2\mathcal A\Gamma}{2\hbar}\left(\frac{\kB\Delta T}{2}\right)^2$ arises when one of the two levels is deep below the equilibrium chemical potential, $-E_\text{L/R}\gg\kBT$ while the other level is located at about $E_\text{R/L}\approx 1.5\kBT$. An explanation for this will be given below.

Similarly to the power, the efficiency is also symmetric under an exchange of the level positions. It takes its maximal value of $\eta\approx 0.1\eta_\text{C}$ in the region $E_\text{L},E_\text{R}>0$ where the output power is strongly suppressed. For these parameters, energy filtering is efficient but the number of electrons that can pass through the filter is exponentially suppressed. For level positions that maximize the output power, the efficiency is slightly reduced to $\eta_\text{maxP}\approx 0.07\eta_\text{C}$.
This efficiency is much smaller than the efficiency at maximum power of a quantum-dot heat engine with couplings small compared to temperature. The latter lets only electrons of a specific energy pass through the quantum dot. Hence, charge and heat currents are proportional to each other. In this tight-coupling limit, the efficiency at maximum power in the linear-response regime is given by $\eta_\text{C}/2$~\cite{van_den_broeck_thermodynamic_2005}. In contrast, the quantum wells transmit electrons of any energy larger than the level position, because any energy larger than the ground state energy can be expressed as $E_\perp + E_z$, where $E_z$ is the $z$-component and $E_\perp$ the perpendicular component of the electron's kinetic energy. Consequently, even high-energy electrons can traverse the barrier, provided most of the energy is in the perpendicular degrees of freedom, and $E_z$ matches the resonant energy. Therefore, they are much less efficient energy filters.

We now aim to understand why the efficiency at maximum power of the quantum-well heat engine is still only about a factor of three less than the efficiency at maximum power of a quantum-dot heat engine with level width of the order of $\kBT$~\cite{nakpathomkun_thermoelectric_2010,jordan_powerful_2013}. The latter configuration has been shown to yield the maximal output power~\cite{jordan_powerful_2013}. To this end, we analyze the situation depicted in \fref{fig:model}. The right quantum well acts as an efficient energy filter because the number of electrons larger than $E_\text{R}$ is exponentially small. The energy filtering at the left quantum well relies on a different mechanism. In order for an electron of energy $E$ to enter the cavity, we need to have $f_\text{L}(E)>0$ such that the reservoir state is occupied. At the same time, we also require $f_\text{C}(E)<1$ such that a free state is available in the cavity. These conditions define an energy window of the order $\kBT$ which explains why the quantum-well heat engine has an efficiency comparable to that of a quantum-dot heat engine with level width $\kBT$.

We now turn to the discussion of an asymmetric system, $a\neq0$. In this case, both the output power and the efficiency are no longer invariant under an exchange of the two level positions. Instead, we now find that power and efficiency are strongly reduced for $E_\text{L}<0$ and $E_\text{R}>0$ if $a>0$ (for $a<0$, the roles of $E_\text{L}$ and $E_\text{R}$ are interchanged). In contrast, for $E_\text{L}>0$ and $E_\text{R}<0$, power and efficiency are even slightly enhanced compared to the symmetric system. This naturally leads to the question of which combination of level positions and coupling asymmetry yields the largest output power. To this end, in \fref{fig:LinearAsymmetry} we plot the power as a function of the asymmetry $a$ and the level position $E_\text{L}$. We find that the maximal power occurs for $a\approx 0.46$ and $E_\text{L}\approx 2\kBT$ while $-E_\text{R}\gg\kBT$. The resulting power is about 20\% larger than for the symmetric setup. At the same time, the efficiency at maximum power is also increased compared to the symmetric system to $\eta\approx0.12\eta_\text{C}$, i.e., it is nearly doubled.
We remark that the maximal efficiency that can be obtained for the asymmetric system is given by $\eta\approx 0.3\eta_\text{C}$. However, similar to the symmetric setup, this occurs in a regime where the output power is highly suppressed.

We now estimate the output power for realistic device parameters. Using $m_\text{eff}=0.067 m_e$, $T=\unit[300]{K}$, $\Gamma=\kB T$ and $a=0.5$, we obtain $P_\text{max}=\unit[0.18]{W/cm^2}$ for a temperature difference $\Delta T=\unit[1]{K}$. Hence, the quantum-well heat engine is nearly twice as powerful as a heat engine based on resonant-tunneling quantum dots~\cite{jordan_powerful_2013}. We remark that materials with higher effective mass yield correspondingly larger output powers. In addition, the quantum-well heat engine offers the advantages of being potentially easier to fabricate. 
As typical level splittings in quantum wells are in the range of $\unit[200-500]{meV}$~\cite{chang_resonant_1974,bonnefoi_resonant_1985}, narrow quantum wells might also be promising candidates for room-temperature applications though leakage phonon heat currents become of relevance then. Finally, we remark on the robustness with respect to fluctuations in the device properties. For the optimal configuration discussed above, fluctuations of $E_\text{R}$ do not have any effect as long as $-E_\text{R}\gg\kBT$. Fluctuations of $E_\text{L}$ by as much as $\kBT$ reduce the output power by about $20\%$ as can be seen in \fref{fig:LinearAsymmetry}. Hence, our device turns out to be rather robust with respect to fluctuations similarly to the quantum-dot based setup in Ref.~\cite{jordan_powerful_2013}.

\subsection{\label{ssec:nonlinear}Nonlinear regime}
\begin{figure}
	\centering\includegraphics[width=.49\textwidth]{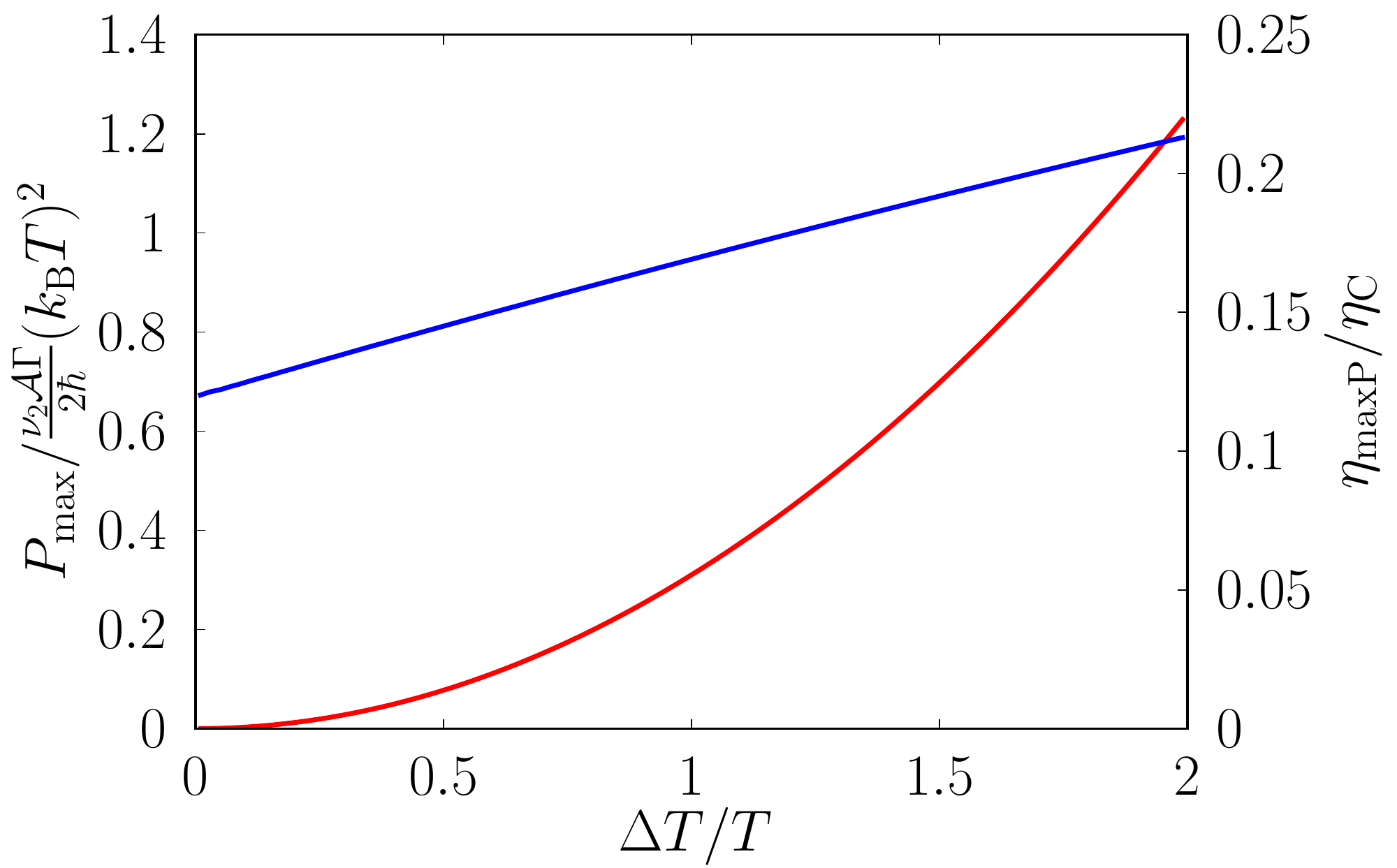}
	\includegraphics[width=.49\textwidth]{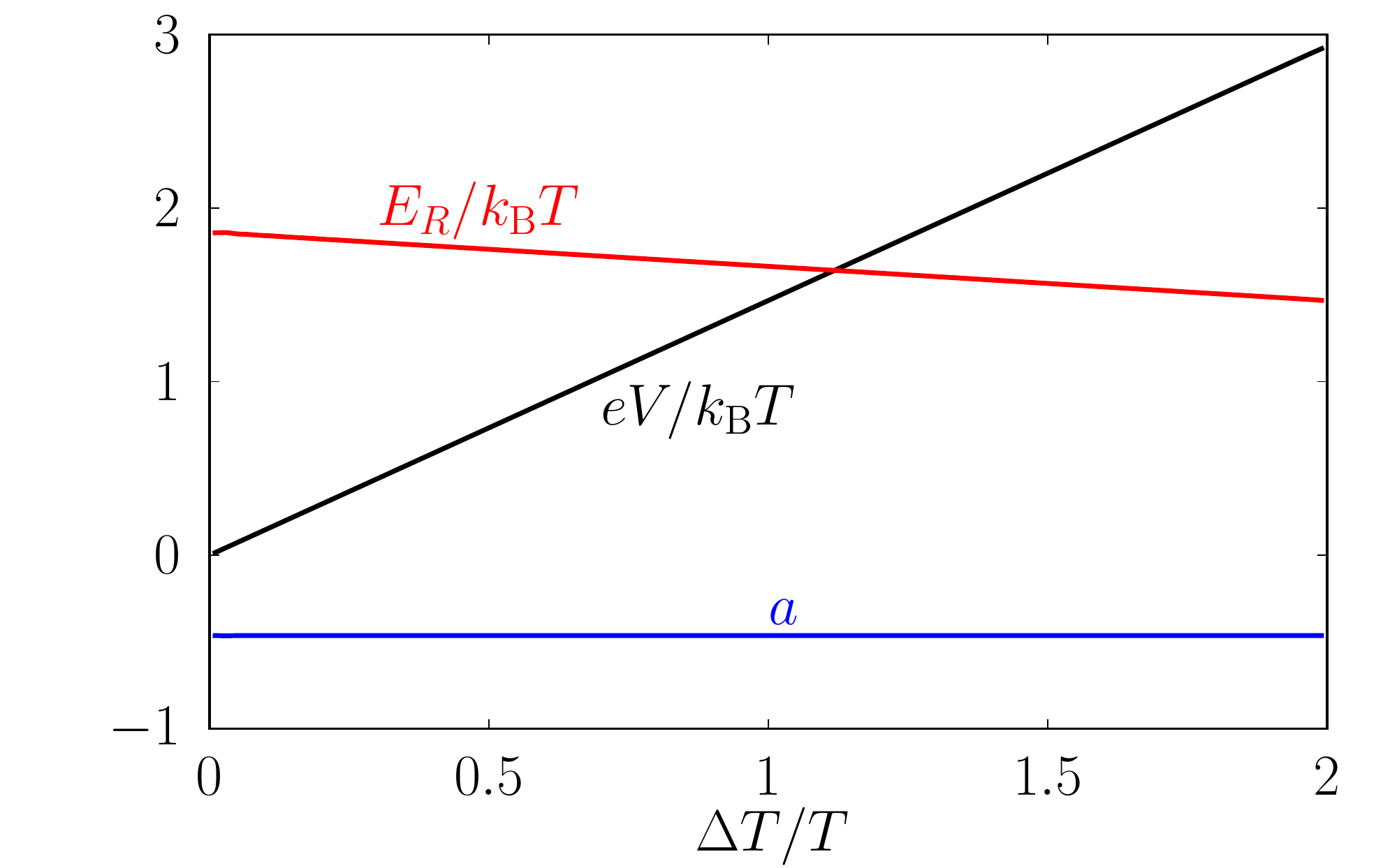}
	\caption{\label{fig:optimal}Left panel: Maximal output power (red) and efficiency at maximum power (blue) as a function of temperature difference $\Delta T$. Right panel: Level position, bias voltage and asymmetry of couplings that maximize the output power as a function of $\Delta T$.}
\end{figure}

We now turn to the performance of the heat engine in the nonlinear regime. Nonlinear thermoelectrics has recently received an increasing interest~\cite{sanchez_scattering_2013,meair_scattering_2013,whitney_thermodynamic_2013}.
We numerically optimized the bias voltage $V$, the asymmetry of couplings $a$ as well as the level positions $E_\text{L,R}$ in order to maximize the output power. The resulting optimized parameters are shown in \fref{fig:optimal} as a function of the temperature difference $\Delta T$. While the optimal asymmetry $a\approx-0.46$ is independent of $\Delta T$, the optimal bias voltage grows linear in $\Delta T$. The right level position $E_\text{R}$ decreases only slightly upon increasing $\Delta T$. The left level position should be chosen as $-E_\text{L}\gg\kBT$, independent of $\Delta T$. It is not shown in \fref{fig:optimal} as our numerical optimization procedure results in large negative values for $E_\text{L}$ that vary randomly from data point to data point because the dependence of the power on $E_\text{L}$ is only very weak in this parameter regime.

The resulting maximal power grows quadratically in the temperature difference, cf. \fref{fig:optimal}. It is approximately given by $P_\text{max}=0.3 \frac{\nu_2\mathcal A\Gamma}{2\hbar}(\kB\Delta T)^2$, independent of $T$. Interestingly, for a given value of $\Delta T$, we obtain the same output power both in the linear and in the nonlinear regime. However, as the efficiency at maximum power grows linearly with the temperature difference, it is preferrable to operate the device as much in the nonlinear regime as possible. In the extreme limit $\Delta T/T=2$, the quantum-well heat engine reaches $\eta_\text{maxP}=0.22\eta_\text{C}$, i.e., it is as efficient as a heat engine based on resonant-tunneling quantum dots while delivering more power~\cite{jordan_powerful_2013}. We remark that the efficiency at maximum power is below the upper bound $\eta_\text{C}/(2-\eta_\text{C})$ given in Ref.~\cite{schmiedl_efficiency_2008}.

\section{\label{sec:conclusions}Conclusions}
We investigated a heat engine based on two resonant quantum wells coupled to a hot cavity. In the linear-response regime we found that our device can yield a power that is nearly twice as large as that of a similar heat engine based on resonant tunneling through quantum dots. At the same time, the efficiency of the quantum-well heat engine is only slightly lower than that of the quantum-dot heat engine. In addition, a device based on quantum wells offers the advantage of being easy to fabricate and the perspective of room-temperature operation. Finally, we also analyzed the performance in the nonlinear regime. There, we found that for a given temperature difference the system yields the same output power as in the linear regime but with an increased efficiency.

In this work, we focussed on the discussion of a setup with noninteracting quantum wells. It is an interesting question for future research how the inclusion of charging effects in the wells relevant in particular in the nonlinear regime affects the performance of quantum-well heat engines.

\ack
We acknowledge financial support from the European STREP project Nanopower, the Spanish MICINN Juan de la  
Cierva program and MAT2011-24331, and ITN Grant No. 234970 (EU). This work was supported by the US NSF Grant No. DMR-0844899.

\appendix
\section*{Appendix}
\setcounter{section}{1}

In this appendix, we give the explicit expression for the function $g_3(x,y)$ that enters the expression \eref{eq:J} for the heat current. It is given by
\begin{eqnarray}
	g_3(x,y)=\frac{2\pi^2}{3}-\frac{1}{2}(x-y)g_1(x,y)\left[x-y-2g_2(x,y)\right]\nonumber\\
	-2(1+a){\rm Li}_2\left(\frac{1}{1+e^{-x}}\right)-2(1-a){\rm Li}_2\left(\frac{1}{1+e^{-y}}\right)\nonumber\\
	-2(1+a)\log(1+e^{x})\log(1+e^{-x})-2(1-a)\log(1+e^{y})\log(1+e^{-y})\nonumber\\
	-g_1(x,y)(1+e^x)(1+e^y)\log(1+e^{-x})\log(1+e^{-y})\nonumber\\
	-g_1(x,y)\log^2(1+e^{-x})\left[e^x\sinh x+\frac{1+a}{1-a}e^x(1+e^y)\right]\nonumber\\
	-g_1(x,y)\log^2(1+e^{-y})\left[e^y\sinh y+\frac{1-a}{1+a}e^y(1+e^x)\right],
\end{eqnarray}
and satisfies the bounds $0<g_3(x,y)<2\pi^2/3$.

\section*{References}


\providecommand{\newblock}{}

\end{document}